\documentstyle[spie]{article} 
\input{psfig}   

\def\la{\hbox{\rlap{\raise.3ex\hbox{$<$}}\lower.8ex\hbox{$\sim$}\ }}
\def\ga{\hbox{\rlap{\raise.3ex\hbox{$>$}}\lower.8ex\hbox{$\sim$}\ }}

\title{Design and Testing of a Prototype Pixellated CZT Detector and
Shield for 
Hard X-Ray Astronomy} 

\author{P. F. Bloser, J. E. Grindlay, T. Narita, 
and J. A. Jenkins
\skiplinehalf 
Harvard-Smithsonian Center for Astrophysics, 60 Garden St., 
Cambridge, MA 02138, USA
}

\authorinfo{Further author information: (Send correspondence to
P. Bloser)\\P.B.: E-mail: pbloser@cfa.harvard.edu}

\begin{document} 
  \maketitle 

\begin{abstract}
We report on the design and laboratory testing of a prototype imaging CZT
detector intended for balloon flight testing in April 2000.  The
detector tests several key techniques needed for the construction of
large-area CZT arrays, as required for proposed hard X-ray astronomy
missions.  Two 10 mm $\times$ 10 mm $\times$ 5 mm CZT detectors, each
with a 4 $\times$ 4 array of
1.9 mm pixels 
on a 2.5 mm pitch, will be mounted in
a ``flip-chip'' fashion on a printed circuit board carrier card; the
detectors will be 
placed 0.3 mm apart in a tiled configuration such that the pixel
pitch is preserved across both crystals.  One detector is eV
Products high-pressure Bridgman CZT, and the other is IMARAD
horizontal Bridgman material.  Both detectors are read out
by a 32-channel VA-TA ASIC controlled by a PC/104 single-board
computer.  A passive shield/collimator surrounded by plastic scintillator
surrounds the detectors on five sides and provides a $\sim 45^{\circ}$
field of view.
The background spectrum recorded by this instrument will be compared
to that measured by a single-element CZT detector (10 mm $\times$ 10
mm $\times$ 2 mm high-pressure Bridgman material from eV Products)
fitted with the same 
passive/plastic collimator but including an active BGO shield to the
rear.
This detector has been previously flown by us completely shielded by a
passive cover.
We describe
preliminary laboratory results for the various systems, discuss
initial background 
simulations, and describe our plans for balloon flight tests.
\end{abstract}



\section{INTRODUCTION}
\label{sec:intro}  

Hard X-ray and gamma-ray detectors made of Cadmium Zinc Telluride (CZT)
hold great promise for advancing the state of X-ray and gamma-ray astronomy
instrumentation.  The intrinsic energy resolution of semiconductor
detectors is far greater than that of scintillators, and the use of
pixel or strip electrode readouts allows far greater spatial
resolution.  The high density of CZT permits the photoelectric
absorption of photons up to $\sim 500$ keV with reasonable thicknesses
($\sim 5$ mm), and the high bandgap allows
detectors to operate at room temperature (as opposed to germanium).
It has been shown that the poor hole transport properties of CZT can
be overcome by special readout electrode geometries that are sensitive
to the motion of electrons only\cite{barrett95}.  We have been
pursuing a program to develop CZT detectors for astronomy
applications, focusing specifically on the needs of a
wide-field-of-view survey telescope operating in the hard X-ray band
between 20 and 600 keV, 
such as the EXIST or EXIST-LITE 
concept\cite{grindlay95,grindlay98}.  

In this paper we describe the CZT instruments we are preparing for a
balloon flight in April 2000 as piggyback experiments on the Harvard
EXITE2 payload.  They are designed to test several of
the key techniques that are needed to construct a
large-area CZT detector plane, and to measure the CZT background at
balloon altitudes with two different shielding configurations under
consideration for a wide-field survey instrument.

\section{TECHNICAL ISSUES IN CONSTRUCTING A HARD X-RAY SURVEY
TELESCOPE}
\label{sec:tech} 

Many technical issues must be addressed before hard X-ray
detectors suitable for astronomy can be constructed.  The only
practical method for imaging between 100 keV and 500 keV is the coded
aperture technique\cite{caroli87}, which requires a large area,
position-sensitive detector.  The wide field of view ($\sim
45^{\circ}$) of an individual survey telescope
module\cite{grindlay95,grindlay98}  together with
the thick detectors 
needed for 
high energy response 
require relatively large pixels (1.5--2.0 mm) to avoid projection
effects.   
Our work so far had thus focused on
thick (5 mm) detectors with 1--2 mm pixels\cite{bloser98mrs}.  At the
same time, the resistivity of the material must be 
large to keep the noise due to leakage current from degrading the energy
resolution, a problem made worse by large pixels.  Thus we have also
investigated the use of blocking contacts made from PIN junctions to
reduce leakage current noise\cite{narita98}.  

The sensitivity intended for EXIST ($<
0.1$ mCrab) requires several square meters of detector material.
Presently the  method of CZT crystal growth yielding the highest
resistivity is the high-pressure
Bridgman (HPB) technique.  Although crystals grown in this manner have high
resistivity, the process is costly since the yield of defect-free
crystals any larger than 10 mm $\times$ 10 mm is low.  A major issue is then
the construction of a large area detector by closely tiling thousands
of small (1 cm$^2$) pixellated crystals with minimal dead space at at
minimum cost, while incorporating the associated readout electronics.
Recently IMARAD Imaging Systems has begun producing CZT using a
modified horizontal Bridgman (HB) process\cite{cheuvart90} which
allows the growth of larger crystals (40 mm $\times$ 40 mm) at higher yield
and thus lower cost.  The drawback is that the crystals have lower
resistivity and thus higher leakage current.  We are investigating
IMARAD CZT detectors using various blocking contacts to reduce this
leakage current noise\cite{narita99}.  

The main technical challenges
are then creating a closely-tiled array out of small detector
elements, reading out the thousands of pixels with the detectors and
electronics in a compact package, and processing the signals from each
channel.  We have already begun working with IDE Corporation to
test low-noise preamps and shaping amps in the form of application
specific integrated circuits (ASICs) that are made as small as
possible\cite{bloser98mrs,narita98}.  The new CZT experiment described
in detail
in Section~\ref{sec:design} begins to meet these challenges by placing
two 10 mm $\times$ 10 mm $\times$ 5 mm CZT detectors with $\sim 2$ mm pixels next to
each other on a common carrier board in such a way that the pixel
pitch is preserved.  The pixels are mounted in ``flip-chip'' fashion
such that the pixels are read out directly into traces on the carrier
board and fed into a 32-channel ASIC controlled by a 
PC/104 single-board computer.  One detector is eV Products HPB
material, the other IMARAD HB material with blocking contacts, which
allows comparitive measures of in-flight background and performance
under very similar conditions.

\section{DESIGN OF THE BACKGROUND EXPERIMENT}
\label{sec:back}

Another critical issue for hard X-ray astronomy instrumentation is the
background level in the detector system.  Astronomical sources 
are faint in the hard X-ray range, and in coded aperture telescopes
the noise per pixel is determined by the total counts in the entire
detector.  To
achieve good signal-to-noise, therefore, the detector background must
be kept to 
a minimum.  Typically the background in balloon payloads is due to a
combination of diffuse cosmic gamma-rays with gamma-ray photons and
energetic particles resulting from cosmic ray interactions in the
atmosphere and in the payload itself.  Effective shielding requires a
detailed knowledge of the 
physical processes that produce background counts in a given detector
material.  These processes must be determined by measurements and
simulations. 
In May 1997 we flew a simple CZT background experiment consisting of a
single-element CZT detector (10 mm $\times$ 10 mm $\times$ 2 mm)
completely shielded with a passive 
Pb/Sn/Cu cup covering the front and sides and actively-shielded by a large BGO
crystal to the rear\cite{bloser98}.  We found that the background rate
in this CZT/BGO detector
was reduced by a factor of $\sim 6$ when BGO triggers were used to
veto events, and that the resulting ``good event'' rate ($9 \times 10^{-4}$
cts cm$^{-2}$ 
s$^{-1}$ keV$^{-1}$ at 100 keV) could be
explained using GEANT simulations that included only gamma-rays
leaking through the shields and produced in the surrounding passive
material.  Gamma-ray interactions alone could not, however, explain
the six-times higher background that was rejected by the BGO,
indicating that an internal activation component was also present that
was effectively vetoed by the active shield.  An
additional goal in flying the new pixellated detectors is to make
another measurement of the flight background spectrum with new shielding
configurations, as well as to study its spatial distribution.

\begin{figure}
\begin{center}
\begin{tabular}{c}
\psfig{figure=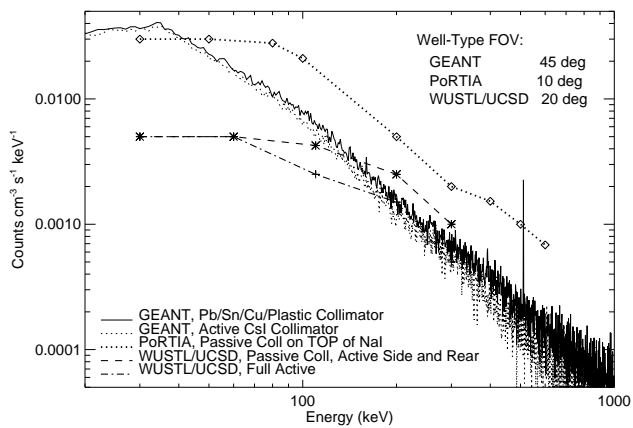,height=6cm} 
\end{tabular}
\end{center}
\caption[initial] 
{ \label{fig:initial}        
GEANT simulations of CZT detectors with active and passive/plastic
collimators and an active rear shield, compared with previous
measurements made using active rear shields and active or passive
collimators.  Spectra are plotted per volume to account for differing
detector thicknesses.  The fields of view are 45$^{\circ}$ for the
GEANT simulations, 10$^{\circ}$ for PoRTIA, and 20$^{\circ}$ for the
WUSTL/UCSD experiment.
} 
\end{figure}

The large reduction in background achieved with the active shield in
the CZT/BGO experiment led
us at first to consider an active collimator of CsI combined with a
rear CsI shield for the current imaging detector experiment.  (BGO was
not considered due to its cost.)  However, detailed simulations have
found that surrounding CZT detectors with thick material leads to
increased background from activation, even if that material is
active, and that a thin passive collimator with an active rear shield is
preferable\cite{armstrong99}.  We investigated the reduction in
gamma-ray-generated background expected from an active collimator by
performing two simple Monte-Carlo 
simulations using the CERN Program Library simulations package GEANT:
a CZT detector at the bottom of a square well 
active
collimator ($45^{\circ}$ field of view) made of 2.5 cm thick CsI, and
a detector at the bottom of a 
square graded passive collimator made of 4 mm of Pb, 1 mm of Sn, and 1
mm of Cu.  In both cases the CZT sat in front of a 2.5 cm thick CsI
rear shield.  Only the interactions of cosmic and atmospheric
gamma-rays were considered (the passive collimator was
assumed to be surrounded by plastic scintillator that vetoed
gamma-rays produced locally from particle interactions).  The spectra
recorded per 
volume of CZT are shown in Figure~\ref{fig:initial}.  The CsI shield
threshold was 50 keV.
The spectrum recorded with the passive collimator is practically
identical to that found with the active collimator
at low energies, 
where aperture flux dominates, and is only $\sim 50$\% higher at
several hundred keV, suggesting that un-vetoed Compton scatters have
only a modest effect on the overall background.  This,
together with the activation simulations mentioned above, indicates
that an active collimator may not be optimal.

The factor of 6 reduction observed by the CZT/BGO
experiment\cite{bloser98} indicates that a large internal background is
being generated in the CZT and vetoed by the BGO.  This background is
not due to shield-leakage gamma-rays, but presumably particle
interactions such as prompt (n,$\gamma$) reactions.  Thus we believe
an active shield must be included in any CZT hard X-ray telescope.
Though it has little effect on the gamma-ray component, would an
active collimator further reduce the internal background component?
In addition to our CZT/BGO balloon flight
experiment, two other CZT background measurements at
balloon altitudes have been performed using an active rear shield and
active or passive collimation, and these have influenced the design of
the present experiment\cite{parsons96,slavis98,slavis99}.  In 1995, a
group from Goddard Space Flight Center flew the CZT experiment PoRTIA,
containing a 25.4 mm $\times$ 25.4 mm $\times$ 1.9 mm planar detector,
in a number of configurations, including passively-collimated to a
$10^{\circ}$ field of view while sitting on top of a thick NaI
crystal\cite{parsons96}.  In 1998 groups from Washington University,
St. Louis (WUSTL) and the University of California-San Diego (UCSD)
flew a 12 mm $\times$ 12 mm $\times$ 2 mm CZT detector with orthagonal strip
electrodes in several different configurations, including one with
active CsI shielding to the rear and sides with a passive collimator
($20^{\circ}$ field of view), and one in which the passive collimator
was replaced with an active NaI collimator\cite{slavis98,slavis99}.
In Figure~\ref{fig:initial} we have included the spectrum recorded by
PoRTIA in the passive collimator/active rear shield, and the spectra
measured by the
WUSTL/UCSD detector in both the active shield/passive 
collimator and fully actively-shielded cases.  All spectra are shown
per volume to allow for differences in detector thickness.  The PoRTIA
background is $\sim 3$ times higher than the GEANT spectrum at all
but the lowest energies, where the larger aperture flux from a
45$^{\circ}$ field of view dominates.  It is not clear why the PoRTIA
background is so much higher in this configuration, but it could be
caused by local gamma-ray production in the passive collimator.  The
WUSTL/UCSD passively-collimated spectrum is slightly higher than the
GEANT prediction  above 150 keV while the actively-collimated spectrum
agrees quite well with it.  The active collimator reduces the
background above 100 keV by a factor of 1.6--2.  Below 100 keV the
higher aperture flux 
assumed in the GEANT spectrum dominates.  (The WUSTL/USCD experiment
also used a depth-sensing technique to reduce background at low
energies\cite{slavis98,slavis99}, although this correction has not been
applied to the spectra shown in Figure~\ref{fig:initial}.)  The
WUSTL/UCSD background was 10 
times higher when both the rear and collimating shields were turned
off.  These results indicate 
that active shielding, together with plastic particle shielding around
passive material, is essential for achieving low backgrounds in
CZT hard X-ray telescopes, and the simple GEANT simulations represent
the ``best-case'' scenario of no internal background.  The rear active
shield does most of the work, however, and the additional reduction
from the active collimator may not be worth the added complexity
and the volume taken up by thick scintillator crystals, especially if
small segments of the CZT array need to be collimated individually.
We therefore decided not to fly an active collimator with our imaging
detectors.  

To measure directly the importance of active shielding for our wide
field of view survey
telescope application, we have decided to fly two simultaneous
background experiments on the next flight of the EXITE2 payload.  The
new pixellated, tiled detectors will be 
flown entirely passively-shielded, with a passive collimator
(45$^{\circ}$ field of view) surrounded by plastic scintillator in the
front and a passive/plastic rear shield in back.  At the same time,
the CZT/BGO detector will be flown again with the
Pb/Sn/Cu cup in front replaced by a
passive/plastic collimator identical to that on the pixellated CZT
detector.  

\section{DESCRIPTION AND TESTING OF INSTRUMENT}
\label{sec:design}

The present experiment tests several of the key elements discussed
in Section~\ref{sec:tech} that are needed for the development of
coded-aperture CZT hard X-ray survey telescopes.  Two thick CZT
detectors (10 mm $\times$ 10 mm $\times$ 5 mm) will be fabricated and
flown in a 
tiled arrangement such that the pixel pitch is preserved across both
detectors.  This is the first important step in building up a large
detector area out of small crystal elements.  One of the detectors
will be eV Products HPB material with gold contacts, similar to
detectors we have tested at length\cite{bloser98mrs}.  The second
detector will be IMARAD HB material made with blocking contacts to
reduce its leakage current and improve energy
resolution\cite{narita99}.  The exact choice of contact material for the
IMARAD detector to be flown has yet to be determined, but for initial
testing we have inserted the IMARAD Au/In detector we have tested
previously in the lab\cite{narita99}.  This detector was manufactured
for us by IMARAD with indium pixels and a gold cathode that acts as a
blocking contact.  Both detectors will have 1.9 mm pixels on a 2.5 mm
pitch (the IMARAD standard pixel size, for compatibility) and will
thus operate within the ``small-pixel regime.''\cite{barrett95}  An
outer guard ring will prevent surface leakage current around the
edges.  This requires making the outer pixels slightly smaller so that
the two detectors may be tiled together while preserving pixel pitch.

\begin{figure}[t] 
\begin{minipage}[t]{3.3in}
\psfig{file=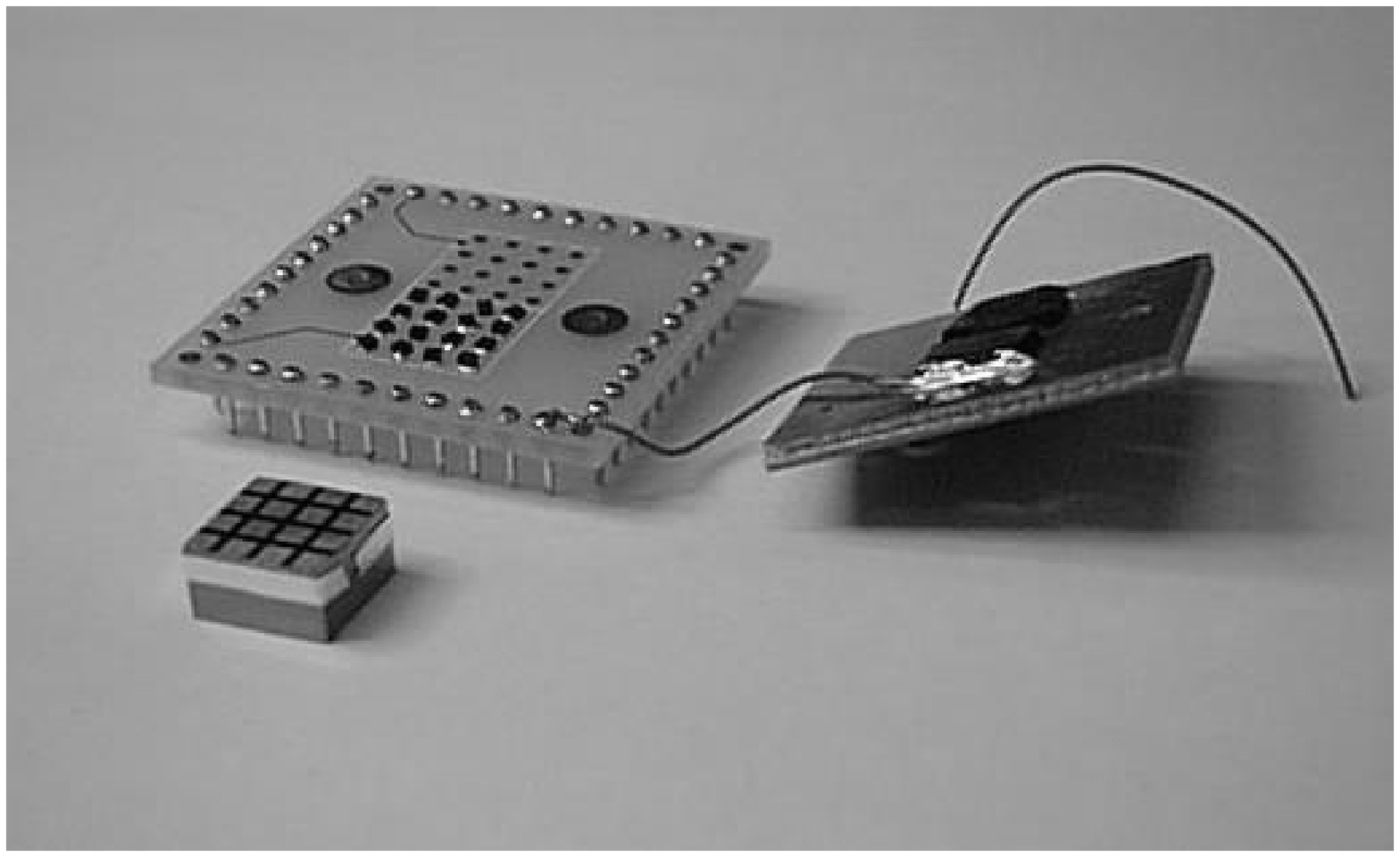,height=2.1in,width=3.2in}
\caption{The IMARAD Au/In detector with the ceramic flip-chip carrier
card.  Conductive rubber pads connect the pixels to the pins via
traces.  The top cover delivers high voltage.}
\label{fig:cztboard}
\end{minipage}
\hspace*{0.2in}
\begin{minipage}[t]{3.3in}
\psfig{file=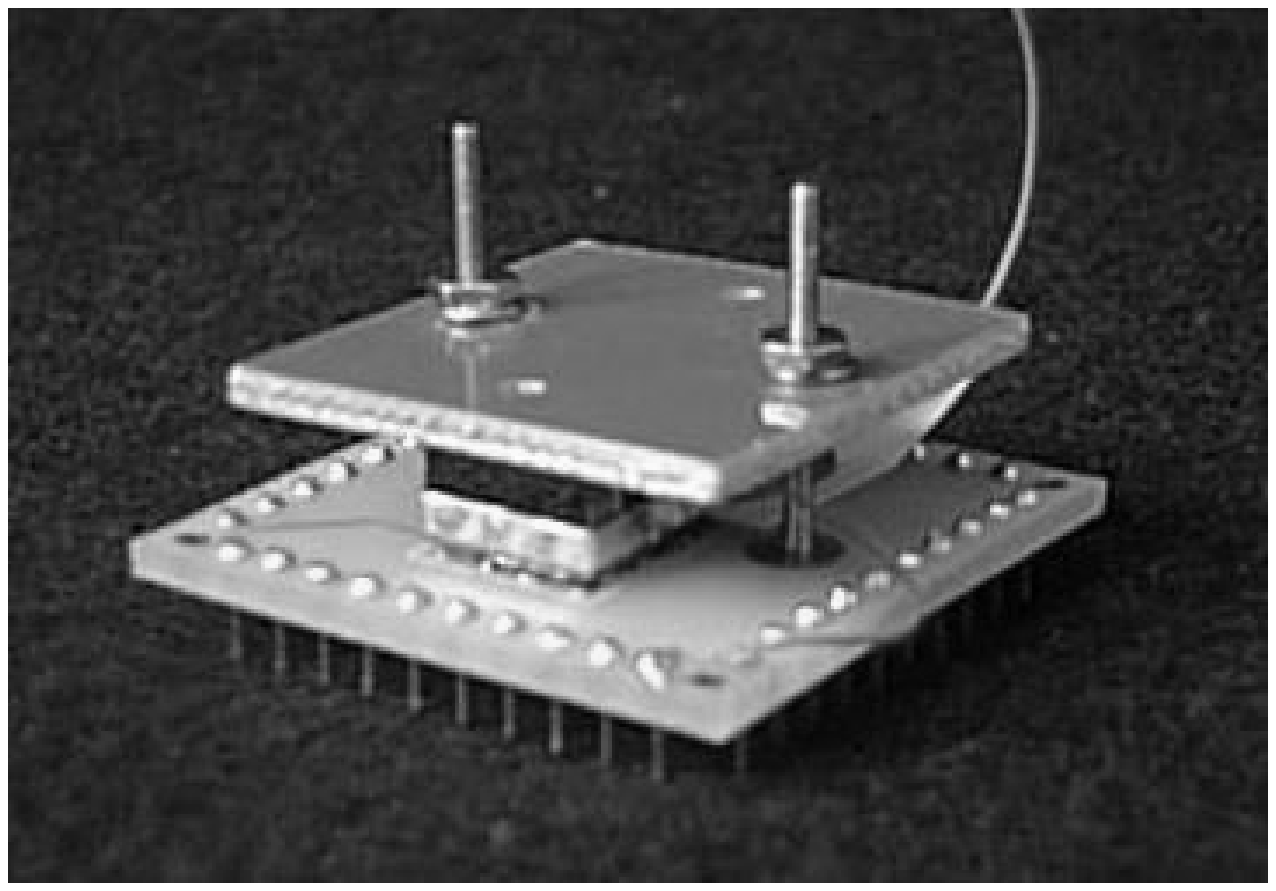,height=2.1in,width=3.2in}
\caption{The assembled flip-chip CZT detector.  The top cover (G10
material) holds 
the CZT between conductive rubber pads for readout and application of
bias voltage.} 
\label{fig:flipchip}
\end{minipage}
\end{figure}

The two pixellated detectors will be mounted in a ``flip-chip'' style
on a specially-designed printed circuit board carrier card (made of
standard FR-4 PCB material).
Figure~\ref{fig:cztboard} shows the IMARAD Au/In detector together
with the flip-chip carrier board and its cover.
Gold pads are arranged to match each pixel and connect it to a pin
via traces on the underside of the board.  To allow us to remove and replace
detectors easily, the electrical connection between the
pixels and traces is made with conductive rubber pads held in place
with a conductive epoxy made by TRA-CON, Inc.  The top cover provides
the negative high voltage to the cathode through another conductive
rubber pad, which is connected by a wire to the bias voltage pin on
the board.  The cover is made of G10 material 0.06'' thick and holds
the detector 
in place between the rubber pads when it is screwed down.
Figure~\ref{fig:flipchip} shows the assembled carrier board with the
IMARAD Au/In detector in place.  We have measured the transmission of
the G10/rubber cover using the low energy lines of a $^{241}$Am
source.  We find transmissions of 6\% at 17 keV (Np line), 60\% at
26.34 keV, and 90\% at 60 keV.  The low energy absorption appears to
be dominated by the G10, so to optimize
the response down to 20 keV we will investigate alternative cover materials.

The two CZT detectors are read out by a 32-channel VA-TA ASIC
manufactured by IDE 
Corp.  The flip-chip carrier card plugs directly into a custom-made
circuit board that contains the ASICs and associated bias resistors
and decoupling capacitors.  The VA-TA combination is attractive
because it includes a self-trigger and MUX to output all 32 channels
for each event.  This will allow us to study the contribution of
multiple-pixel events to the background and possibly to correct for the
effects of charge-spreading between adjacent pixels\cite{narita98}.
The expected count rate is $\sim 1$ count cm$^{-2}$ s$^{-1}$ (see
Figure~\ref{fig:expected}), or $\sim 2$ counts s$^{-1}$ from all 32
pixels together, easily low enough to record all channels for each
event.  The VA-TA ASICs are controlled by a data acquisition (DAQ)
board supplied by IDE.  We are in the process of writing software to
control this DAQ board that will run on a PC/104 single-board computer
flown alongside the detectors.  This computer will record data into
buffers and
transfer them, along with housekeeping data, into the main EXITE2
data stream.  

\begin{figure}[t] 
\begin{minipage}[t]{3.3in}
\psfig{file=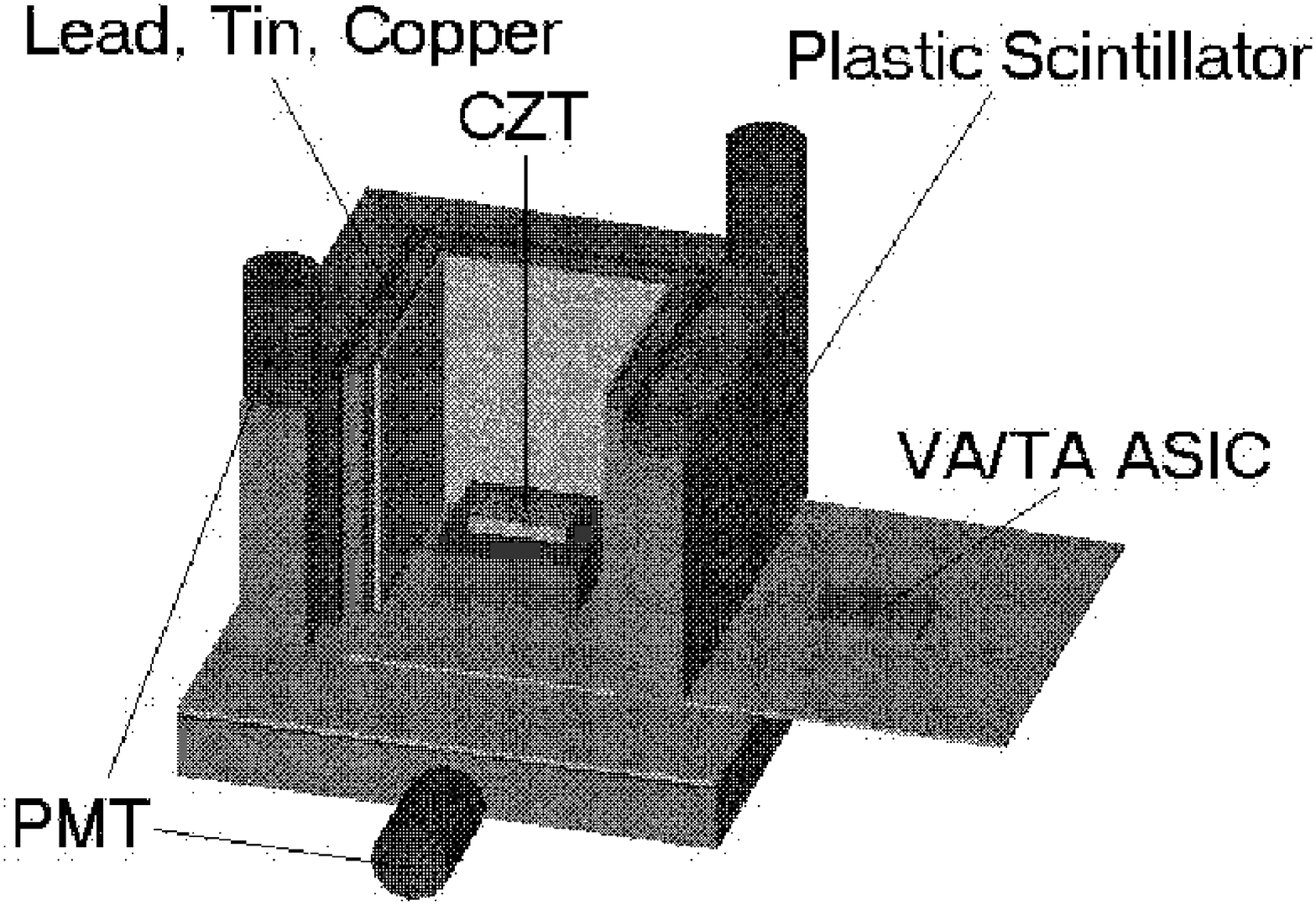,height=2.1in,width=3.2in}
\caption{Layout of the tiled CZT experiment shielding, showing the
detectors, VA-TA board, Pb/Sn/Cu collimator and rear shield, NE-102
plastic scintillator particle shield, and readout PMTs.}
\label{fig:shield2}
\end{minipage}
\hspace*{0.2in}
\begin{minipage}[t]{3.3in}
\psfig{file=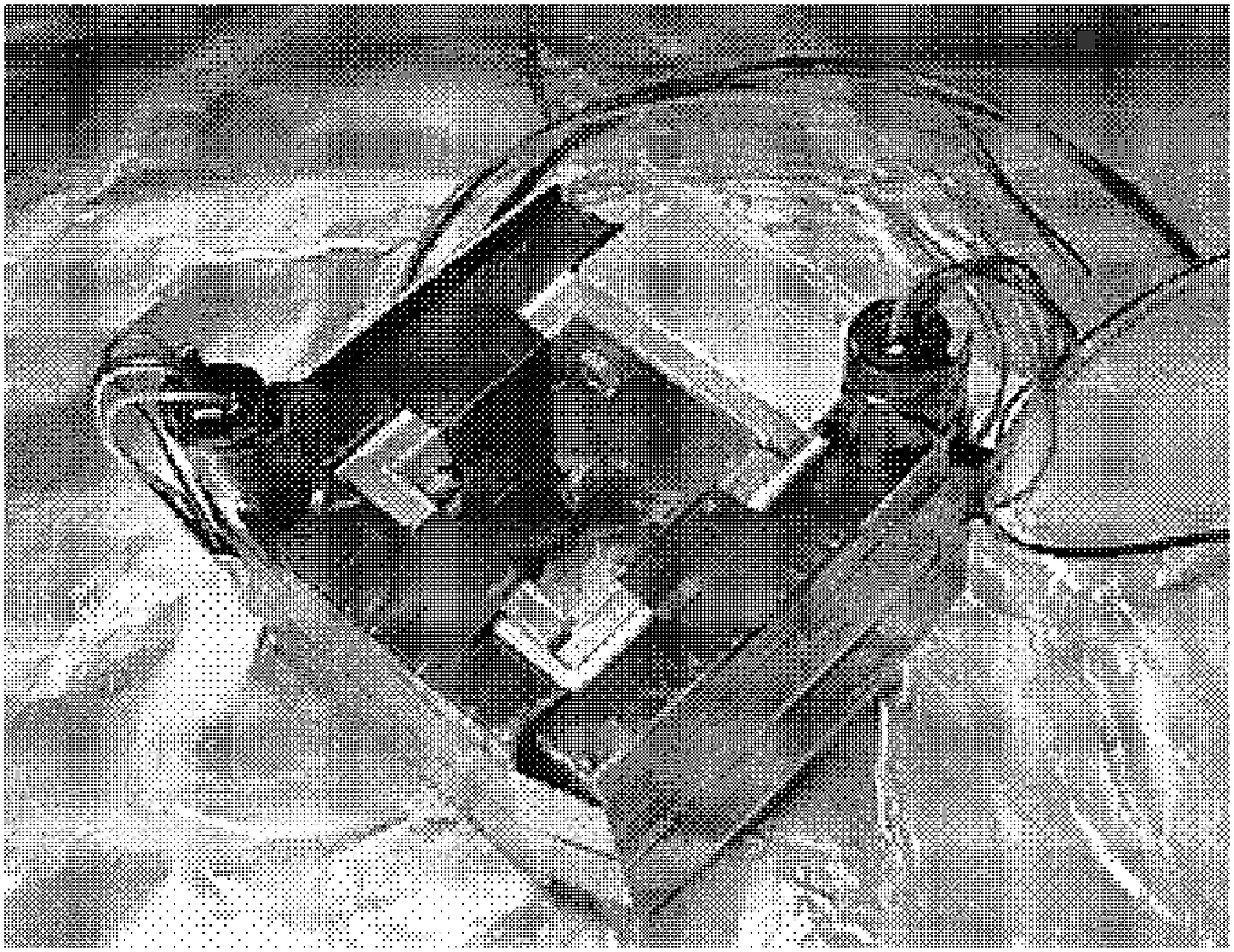,height=2.1in,width=3in}
\caption{The assembled passive/plastic collimator.  The Pb, Sn, and Cu slats
are bolted on the inside of an aluminum frame, while the plastic
shield takes the form of two L-shaped halves, each read out in the
corner by a miniature PMT.} 
\label{fig:realshield}
\end{minipage}
\end{figure}

As described in Section~\ref{sec:back}, the two tiled flip-chip
detectors will be surrounded by a passive/plastic collimator and rear
shield.  This is shown schematically in Figure~\ref{fig:shield2}.
In order to keep the instrument weight down, the collimator only
surrounds the CZT carrier board ($\sim 3$ cm across), providing a
$45^{\circ}$ field of 
view.  Since the current prototype VA-TA board is too large to fit
within this 
space, it was necessary to have the collimator and rear shield
physically separated.  This required making the rear shield large
enough to prevent the detectors from having a line of sight to the
outside.  Future designs will stipulate that the ASIC and its circuit
board fit entirely within the footprint of the detector carrier card,
so that they may be assembled vertically and fit within the main
shielding.

Figure~\ref{fig:realshield} shows the assembled passive/plastic
collimator.  The passive portion consists of 4.5 mm Pb, 1 mm Sn, and 1
mm Cu slats bolted within an aluminum support frame.  
Cosmic ray particles will interact in this dense material, generating
gamma-ray photons.  To prevent these locally-produced gamma-rays from
producing background in the CZT, plastic shields surround the passive
material to provide a veto pulse when charged particles pass through
them.  The shields are made of 0.5'' thick NE-102 plastic scintillator
joined with Bicron BC-600 optical cement into two L-shaped halves.
The readout devices for such shields (and active gamma-ray shields as
well) must be extremely compact if large-area arrays are to be built
up of smaller detector/shield units.  We have selected Hamamatsu
R7400U miniature photomultiplier tubes (PMTs) as the readout devices
for the plastic shields.  These PMTs are only 1.5 cm across and 2.6 cm
long, have far higher gain than photodiodes or avalanche photodiodes
(APDs), and their gain is not 
temperature dependent as in APDs.  One PMT is placed
in the corner of each L to read out two sides of the plastic shield.
We have tested this arrangement in the lab by setting up a muon
telescope: two lab PMTs coupled to plastic scintillators were placed
on either side of the shield.  Only muons
passing through all three scintillators generated a coincident signal,
and using this coincidence we could identify pulses from particles
interacting anywhere along the L.  We found that particles passing
through the shield on the opposite end from the readout PMT still
produce easily-measurable pulses, and so the two miniature PMTs are
able to read 
out the entire volume of plastic.  Pulses from the PMTs are fed into a
coincidence-logic card that generates a veto pulse for coincident CZT and
plastic triggers.  The PC/104 computer will recognize this veto
pulse and flag the event. 

\begin{figure}[t] 
\begin{minipage}[t]{3.3in}
\psfig{file=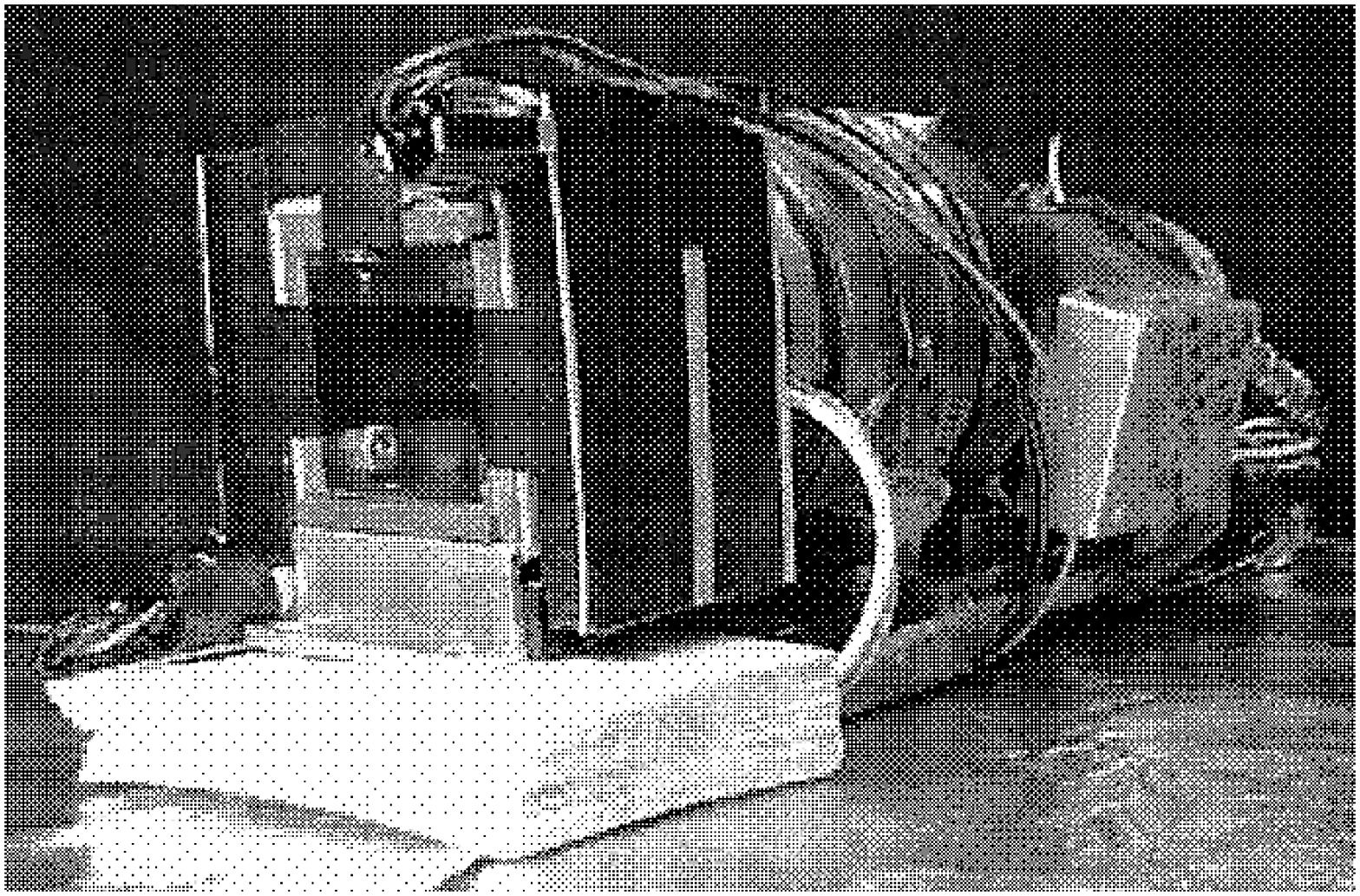,height=2.1in,width=3in}
\caption{The passive/plastic collimator with the CZT/BGO detector in
flight configuration.  The
BGO crystal is housed in the cylindrical container visible to the right
of the collimator, and is read out by the large PMT behind that. This
experiment will test the passive/plastic collimator and active rear
shield configuration.} 
\label{fig:onbgo}
\end{minipage}
\hspace*{0.2in}
\begin{minipage}[t]{3.3in}
\hspace{0.4in}\psfig{file=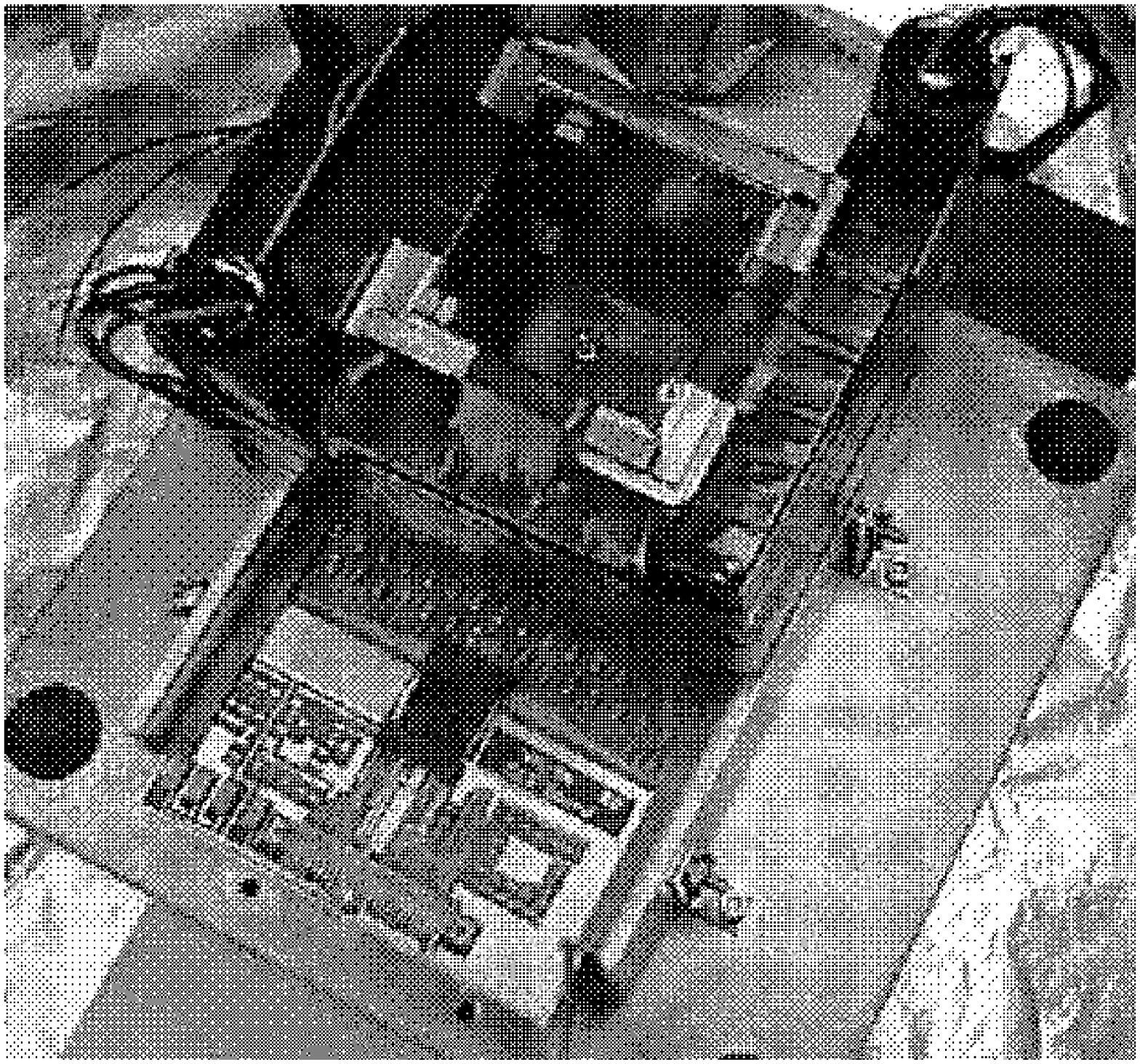,height=2.1in,width=2.5in}
\caption{The passive/plastic collimator shown in the flight
configuration with the tiled flip-chip detector assembly.  The
collimator surrounds the detector carrier board, which is plugged into
the VA-TA board.  The ASICs are mounted under the black rectangular
cover to the lower left.  The rear passive/plastic shield (not shown) will lie
directly beneath the VA-TA board.} 
\label{fig:onnew}
\end{minipage}
\end{figure}

As described in Section~\ref{sec:back}, identical passive/plastic
collimators will be flown with the old single-element CZT/BGO
detector and the new pixellated tiled detectors.
Figure~\ref{fig:onbgo} shows the passive/plastic collimator mounted in front
of the CZT/BGO experiment.  
The CZT detector sits at the rear
of the collimator and observes a $45^{\circ}$ field of 
view.  The BGO crystal is housed within the cylindrical container
directly behind the CZT, and is read out by the large PMT
at the rear.  This setup will test the passive collimator/active rear
shield configuration under consideration for future CZT telescopes.

Figure~\ref{fig:onnew} shows the passive/plastic collimator in the
flight configuration with the flip-chip detector assembly and VA-TA
board.  The detector is mounted in the carrier card, visible at the
bottom of the collimator.  As shown schematically in
Figure~\ref{fig:shield2}, the VA-TA board extends out from under the
collimator as presently constructed.  The large passive/plastic shield
will be mounted directly under the VA-TA board in flight.  This setup
will test the completely passively-shielded configuration for direct
comparison with the actively-shielded case described above.  The ASICs
are mounted under the black rectangular cover visible beneath the
collimator, and the connector that leads to the DAQ board is shown at the
bottom of the figure.

\begin{figure}
\begin{center}
\begin{tabular}{c}
\psfig{figure=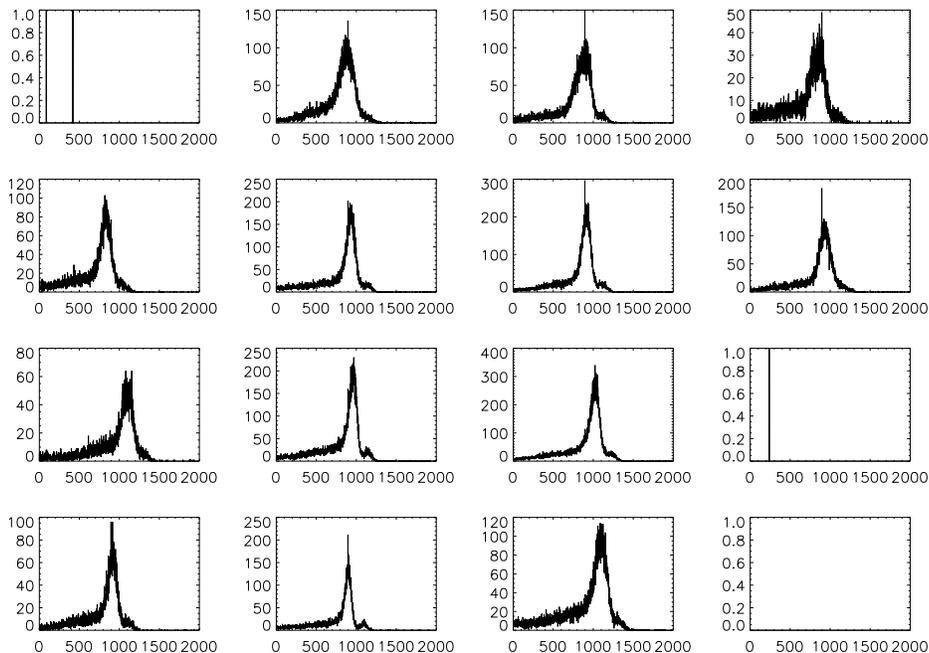,height=9cm} 
\end{tabular}
\end{center}
\caption[initial] 
{ \label{fig:spec}        
``First light'' 4 $\times$ 4 array of $^{57}$Co spectra taken with the IMARAD
Au/In detector 
mounted on the flip-chip carrier card read out by the VA-TA ASIC.
Three pixels were found to be disconnected because the conductive
rubber pads had become detached.  The pixels in the top row are
unusually noisy, either due to bad contact with the pixels or noise in
certain ASIC channels.
} 
\end{figure}

Figure~\ref{fig:spec} shows the 4 $\times$ 4 array of ``first light'' $^{57}$Co
spectra from the 16 pixels of the IMARAD Au/In detector mounted on the
flip-chip 
carrier card and read out through the VA-TA ASIC.  
Three channels are disconnected; it was found that these three conductive
rubber pads had become detached during our initial attempts at
mounting the detector.  In addition, the channels in the top row
are unusually noisy.  Either a good connection was not made between
the pixels and rubber pads, or these channels of the ASIC are picking
up excess noise.  In any case, the spectra in the lower left portion
of the detector are comparable to those taken through more
conventionally-mounted detectors\cite{narita98,narita99}, and so prove
that our flip-chip 
mounting scheme is feasible.

\section{DISCUSSION AND CONCLUSIONS}
\label{sec:disc}

\begin{figure}[t] 
\begin{minipage}[t]{3.3in}
\psfig{file=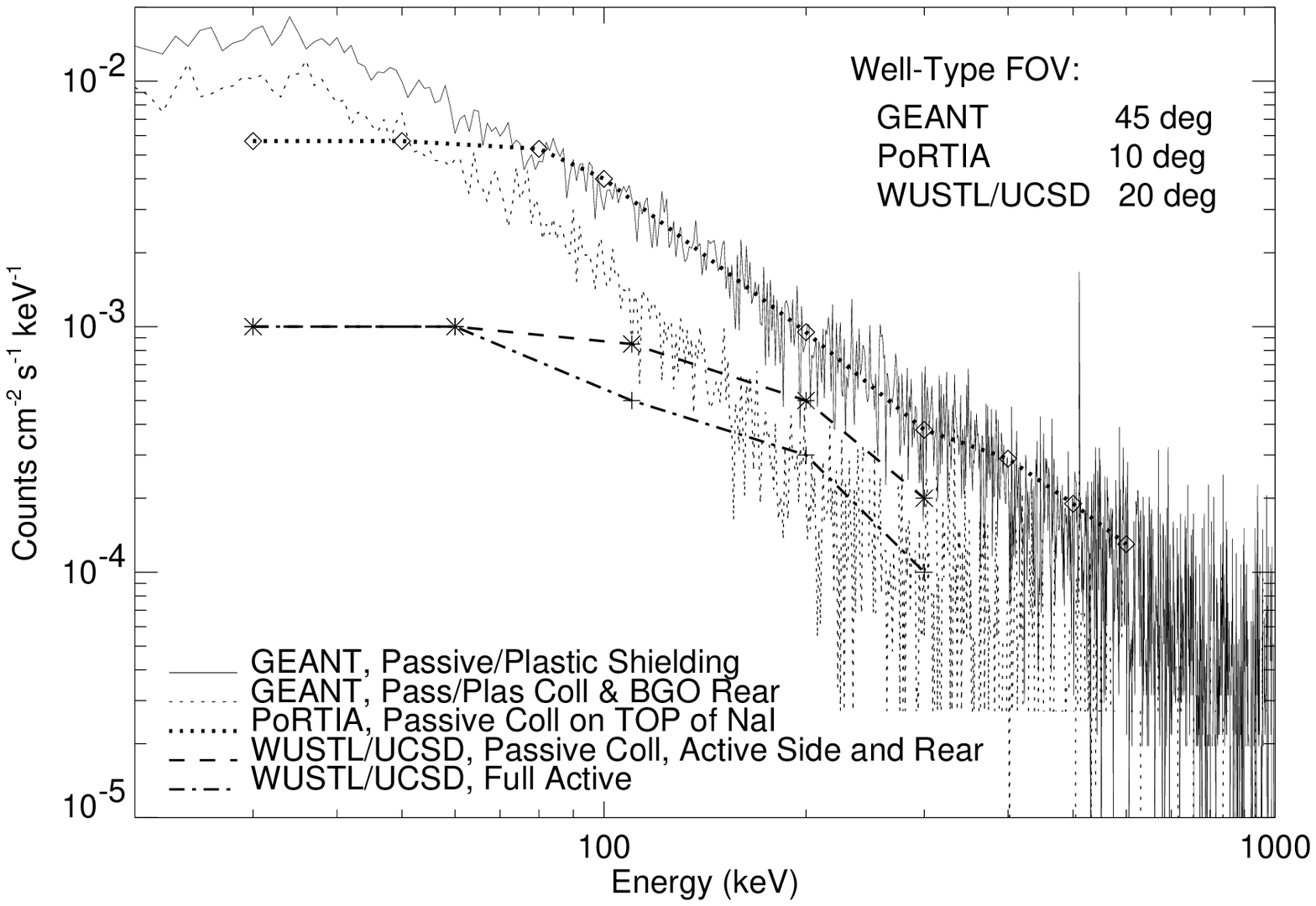,height=2.1in,width=3.2in}
\caption{Expected gamma-ray backgrounds per area expected for the CZT/BGO
detector with
passive/plastic collimator and pixellated
flip-chip detectors with  passive/plastic shielding, as computed by
GEANT.  The PoRTIA 
and WUSTL/UCSD results are reproduced as in Figure~\ref{fig:initial}.
The thicker pixellated detectors record a higher background per area.}
\label{fig:expected}
\end{minipage}
\hspace*{0.2in}
\begin{minipage}[t]{3.3in}
\psfig{file=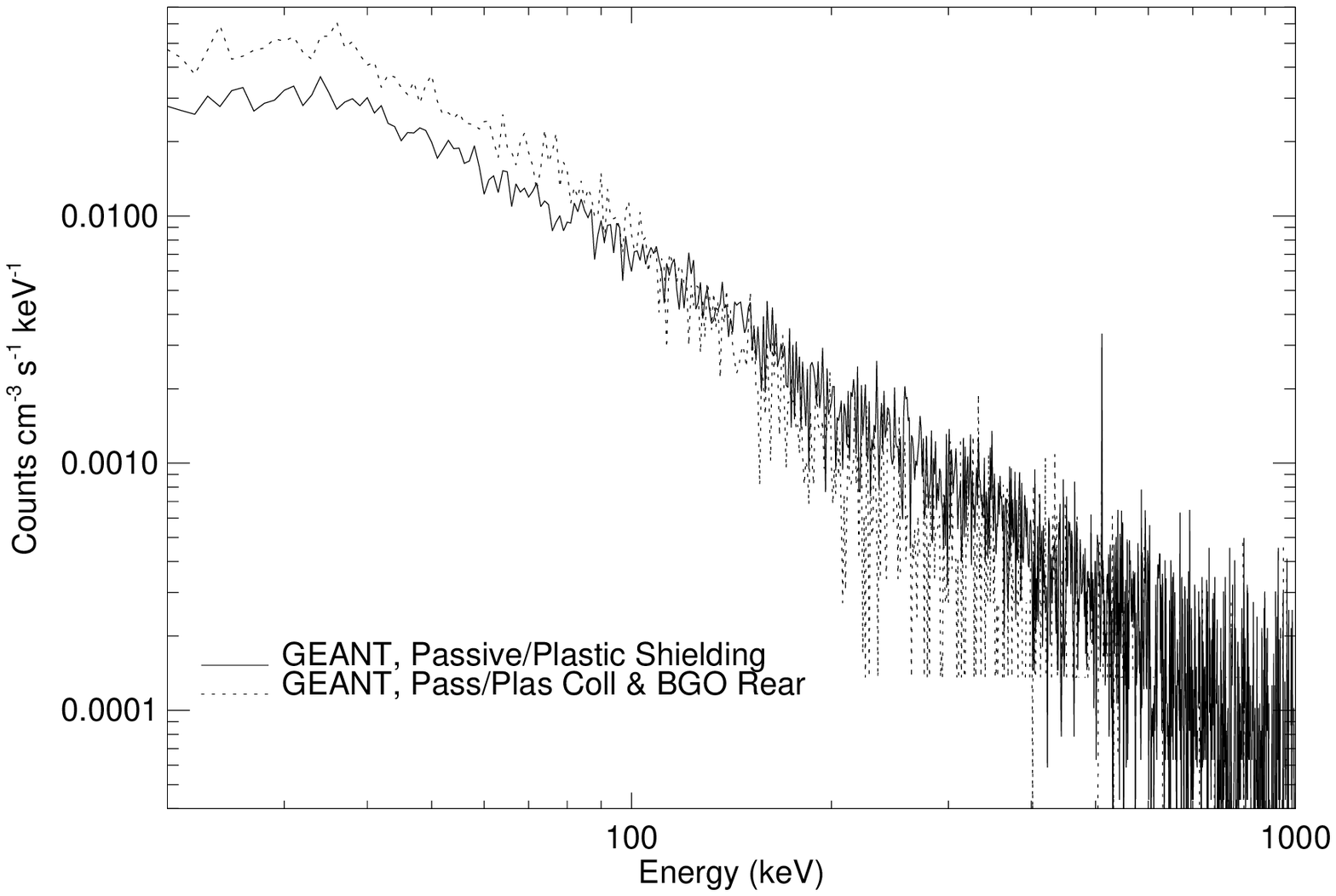,height=2.1in,width=3.2in}
\caption{The GEANT spectra of Figure~\ref{fig:expected} plotted per
detector volume.  At low energies the thinner CZT/BGO detector has
a greater fraction of its volume filled by aperture flux counts, and
so the per volume count rate is higher.  At high energies the CZT/BGO
background is slightly lower, due to rejected Compton scatter events.} 
\label{fig:volume}
\end{minipage}
\end{figure}

Both the CZT/BGO detector with its passive collimator and the
pixellated flip-chip 
detectors with their passive/plastic shielding will be flown as
piggyback experiments on the next flight of the EXITE2 hard X-ray
telescope payload.  This flight is scheduled to take place in April of
2000 from Ft. Sumner, NM.  
The expected gamma-ray contributions to the backgrounds of the CZT/BGO
and flip-chip CZT detectors have been calculated using GEANT in the
same manner as the spectra shown in Figure~\ref{fig:initial}. These
simulations then represent a ``lower limit'' to the total background
expected.  In
Figure~\ref{fig:expected} we 
compare them to the PoRTIA and WUSTL/UCSD results as before.
Here we plot the backgrounds per area, since the important quantity is
the background counts within the area of a pixel.  It is obvious that the
thicker pixellated detectors in the passive/plastic shield can expect to
record a higher background per pixel than would the thinner detectors
in the other experiments.  The pixellated detector background is
higher by a factor of $\sim 1.5$ even at the lowest energies because
these detectors are mounted slightly closer to the front of the
collimator than is the CZT/BGO detector, giving them a slightly larger
field of view and higher aperture flux.  There is good agreement between the
expected CZT/BGO background and the completely actively-shielded
WUSTL/UCSD background at high energies.  This might indicate that the
WUSTL/UCSD experiment has successfully rejected most of the internal
background and is recording mostly shield leakage
photons.  Whether the CZT/BGO
background will really be this low will depend on the efficiency of the
BGO active shield in rejecting prompt internal background.  Our
previous results\cite{bloser98} indicate that the actual spectrum
should lie within a factor of two of that plotted in
Figure~\ref{fig:expected}. 

To compare the CZT/BGO and pixellated experiments without regard to
detector thickness, we plot the GEANT spectra as a function of detector
volume in Figure~\ref{fig:volume}.  The two backgrounds are in fairly close
agreement.  At low energies the background is dominated by aperture
flux from the front.  Both the thin CZT/BGO and thick pixellated detectors
efficiently absorb these low energy photons, but the thin detector
does so in a smaller total volume.  Therefore the count rate in the
thin detector appears higher when plotted per volume.  At high
energies the CZT/BGO 
background is slightly lower due to the rejection of Compton-scattered
events.  In reality, our results\cite{bloser98}, together with those
of the WUSTL/UCSD experiment\cite{slavis98,slavis99}, indicate that
the background in the passive/plastic-shielded detectors will be 6--10
times higher than that in Figures~\ref{fig:expected} or~\ref{fig:volume}.
It will be of great value to measure how much
higher it is, and to attempt to model the processes responsible for it.

As noted in Section~\ref{sec:back}, the WUSTL/UCSD experiment
makes use of a 
depth-sensing technique to further lower the background below 100 keV
by rejecting low energy events near the bottom of the
detector\cite{slavis98,slavis99}.  The method is based on the fact
that the ratio of the cathode and pixel pulses for a given event
should be proportional to the depth of the interaction.  Such a
technique would be even more 
useful in 5 mm thick detectors such as ours, and we will 
implement it in our application by adding an extra channel to future
ASICs to read 
out the cathode pulse.

To fully understand the background measurements we will make, it will
be necessary to model the response of the CZT detectors themselves.
We have already successfully simulated the simple response of the
single-element CZT/BGO detector\cite{bloser98}.  Modeling a pixellated
detector requires knowledge of the internal electric field and
weighting potentials\cite{barrett95}.  We have 
already developed electric field modeling tools based on the
commercial software package ES4, and are now modeling
the response of our tiled
imaging detectors to laboratory X-ray sources.
We will also attempt to include cosmic ray interactions and activation
precesses (e.g. $^{110}$Cd(n,$\gamma$)) in our
background simulations in order to understand the internal background
processes important in CZT.

\acknowledgments     
 
We thank F. Harrison and B. Matthews for providing the CZT/BGO
detector for further flights.  This work was supported in
part by NASA grant NAG5-5103.  P. Bloser acknowledges support from
NASA GSRP grant NGT5-50020.


  \bibliography{newczt_paper}   

\begin{thebibliography}{10}

\bibitem{barrett95}
H.~Barrett, J.~Eskin, and H.~Barber, ``Charge transport in arrays of
  semiconductor gamma-ray detectors,'' {\em Phys. Rev. Lett.} {\bf 75}, p.~156,
  1995.

\bibitem{grindlay95}
J.~Grindlay, T.~Prince, N.~Gehrels, J.~Tueller, C.~Hailey, B.~Ramsey,
  M.~Weisskipf, P.~Ubertini, and G.~Skinner, ``{E}nergetic {X}-ray {I}maging
  {S}urvey {T}elescope ({EXIST}),'' {\em Proc. SPIE} {\bf 2518}, p.~202, 1995.

\bibitem{grindlay98}
J.~Grindlay, ``Balloon-borne hard x-ray imaging and future surveys,'' {\em Adv.
  Space Res.} {\bf 21}, p.~999, 1998.

\bibitem{caroli87}
E.~Caroli, J.~Stephen, G.~DiCocco, L.~Natalucci, and A.~Spizzichino, ``Coded
  aperture imaging in x- and gamma-ray astronomy,'' {\em Space Sci. Rev.} {\bf
  45}, p.~349, 1987.

\bibitem{bloser98mrs}
P.~Bloser, T.~Narita, J.~Grindlay, and K.~Shah, ``Prototype imaging
  {C}d-{Z}n-{T}e array detector,'' in {\em Semiconductors for Room-Temperature
  Radiation Detector Applications II},  R.~James, T.~Schlesinger, P.~Siffert,
  M.~Cuzin, M.~Squillante, and W.~Dusi, eds., {\em Proc. MRS} {\bf 487},
  p.~153, 1998.

\bibitem{narita98}
T.~Narita, P.~Bloser, J.~Grindlay, R.~Sudharsanan, C.~Reiche, and C.~Stenstrom,
  ``Development of prototype pixellated {PIN} {C}d{Z}n{T}e detectors,'' in {\em
  Hard x-ray and gamma-ray detector physics and applications},  {\em Proc.
  SPIE} {\bf 3446}, p.~218, 1998.

\bibitem{cheuvart90}
P.~Cheuvart, U.~El-Hanany, D.~Schneider, and R.~Triboulet, ``{C}d{T}e and
  {C}d{Z}n{T}e crystal growth by horizontal {B}ridgman technique,'' {\em J.
  Crystal Growth} {\bf 101}, p.~270, 1990.

\bibitem{narita99}
T.~Narita, P.~F. Bloser, J.~E. Grindlay, J.~A. Jenkins, and H.~W. Yao,
  ``Development of {IMARAD} {CZT} detectors with {PIN} contacts,'' {\em Proc.
  SPIE} {\bf 3768}, 1999.

\bibitem{bloser98}
P.~Bloser, J.~Grindlay, T.~Narita, and F.~Harrison, ``{C}d{Z}n{T}e background
  measurement at balloon altitudes with an active {BGO} shield,'' in {\em EUV,
  X-ray, and Gamma-ray Instrumentation for Astronomy},  O.~H.~W. Siegmund and
  M.~Gummin, eds., {\em Proc. SPIE} {\bf 3445}, p.~186, 1998.

\bibitem{armstrong99}
T.~W. Armstrong, B.~L. Colborn, and B.~D. Ramsey, ``Initial estimates of
  radiation backgrounds in the cadmium-zinc-telluride focal plane detectors,''
  {\em Science Applications International Corporation Report} {\bf
  SAIC-TN-99015R}, 1999.

\bibitem{parsons96}
A.~Parsons, S.~Barthelmy, L.~Bartlett, F.~Birsa, N.~Gehrels, J.~Naya, J.~Odom,
  S.~Singh, C.~Stahle, J.~Tueller, and B.~Teegarden, ``{C}d{Z}n{T}e background
  measurements at balloon altitudes,'' {\em Proc. SPIE} {\bf 2806}, p.~432,
  1996.

\bibitem{slavis98}
K.~Slavis, P.~Dowkontt, F.~Duttweiler, J.~Epstein, P.~Hink, G.~Huszar,
  P.~Leblanc, J.~Matteson, R.~Skelton, and E.~Stephan, ``High altitude balloon
  flight of {C}d{Z}n{T}e detectors for high energy x-ray astronomy,'' in {\em
  EUV, X-ray, and Gamma-ray Instrumentation for Astronomy},  O.~H.~W. Siegmund
  and M.~Gummin, eds., {\em Proc. SPIE} {\bf 3445}, p.~169, 1998.

\bibitem{slavis99}
K.~R. {Slavis}, P.~F. {Dowkontt}, J.~W. {Epstein}, P.~L. {Hink}, J.~L.
  {Matteson}, F.~{Duttweiler}, G.~L. {Huszar}, P.~C. {Leblanc}, R.~T.
  {Skelton}, and E.~A. {Stephan}, ``Background studies in {CZT} detectors at
  balloon altitudes,'' {\em American Astronomical Society Meeting} {\bf 193},
  pp.~6605+, Dec. 1998.

\end{thebibliography}
  \bibliographystyle{spiebib}   
 
  \end{document}